# 调制稳频对电磁诱导透明的光谱噪声影响*

侯婧华 [1]，苏楠 [2]，刘瑶 [2]，刘智慧 [2]，张玉驰 [1*]，何军 [2**]

[1] 山西大学物理电子工程学院，山西 太原 030006；

[2] 山西大学光电研究所量子光学与光量子器件国家重点实验室，山西 太原 030006

**摘 要** 通过探测光输出噪声谱研究了铯原子阶梯型三能级系统的电磁诱导透明光谱中耦合光的调制信号向探测光的转化以及耦合光额外相位噪声向探测光振幅噪声的转化，证明了外调制稳频可有效避免调制噪声转移。实验中分别采用外调制转移光谱与内调制转移光谱两种稳频方法，实现了耦合光频率稳定，并测量了两种方法下稳频输出端与检测输出端的探测光噪声谱。实验表明在电磁诱导透明介质中耦合光的调制信号可以向探测光转化，外调制稳频引入的调制信号不会影响检测输出端，内调制的检测输出端在调制幅度为 0.004 V 时被引入了约 20 dB 的调制信号；外调制稳频引入的额外相位噪声不会影响检测输出端，内调制稳频引入的额外相位噪声转化为检测输出端的振幅噪声，转化最大处达到约 39 dB。此外，我们测量了外调制稳频输出端、内调制稳频输出端与内调制检测输出端相位噪声大小随调制幅度的变化。结果表明，耦合光相位噪声转化随调制幅度的增加显著增大。

**关键词** 噪声谱；电磁诱导透明；激光频率稳定；额外相位噪声

**中图分类号** O562 **文献标志码** A

# 1 引 言

电磁诱导透明（EIT）效应是典型的非线性相干光学现象。在三能级原子系统中，强耦合光与弱探测光共同作用，通过量子干涉抑制原子对探测光的吸收，使原本不透明的介质呈现透明特性。1988 年 Kocharovslaya 等以及 1989 年 Harris 各自独立地提出了无粒子数反转光放大的概念，为 EIT 的发现奠定早期的理论基础[1-2]。1991 年，EIT 效应由 Harri 研究组首次通过实验观察到[3]。自此以后，EIT 效应不仅在原子气室中得以实现，在冷原子[4]、固态材料[5-7]以及半导体[8]等介质中也被进行了深入地研究。在原子气室或冷原子中采用 EIT 技术，使得基于里德堡原子的微波电场测量[9-12]与通信[13-14]成为了可能，并推动这些领域实现了快速发展。

在基于原子气室的测量与通信系统中，噪声是影响测量灵敏度及通信误码率的关键因素之一，因此人们对 EIT 信号中的噪声特性展开了广泛的研究。2006 年，Hsu 等通过实验发现 EIT 系统的探测光中引入了额外噪声，这是当时理论尚未预测到的[15]。2007 年，Zhang 等通过理论计算发现 EIT 系统中探测光的输出振幅噪声由探测光的振幅噪声、相位噪声以及原子噪声共同决定[16]。2009 年，Barberis-Blostein 等理论计算了 EIT 系统中探测光与耦合光之间存在噪声的振荡交换[17]。2012 年，Li 等通过实验验证了 EIT 系统中探测光和耦合光的额外相位噪声皆可向探测光的振幅噪声转化[18]。2018 年，Jia 等也在实验上证实了耦合光的额外相位噪声可以向探测光的振幅噪声转化[19]。因此 EIT 系统中探测光的振幅噪声、探测光的相

位噪声、耦合光的振幅噪声、耦合光的相位噪声及原子噪声共同决定了探测光的输出振幅噪声。为了减小输出探测光噪声，人们从多方面展开工作。Bai 等利用声光频移器的布拉格衍射方式对激光功率进行控制，改善了激光的振幅噪声[20]。Sun 等通过本地振荡光和耦合光之间的锁相环路有效抑制了实验测量过程中 68.05%的随机相位抖动，改善了激光的相位噪声[21]。Xiao 等展示并描述了两种可减轻 EIT 中激光相位噪声影响的相干现象：一种是 EIT 光场之间透射强度互相关中抗激光功率展宽的共振，另一种是当单光子噪声主导双光子失谐噪声时，激光相位噪声到振幅噪声转换的共振抑制[22]。He 等提出当探测光和耦合光的频率满足双光子共振条件时，在零失谐处很窄的范围内，相位转换被强烈抑制[23]。以上抑制噪声的工作主要集中在激光功率起伏产生的入射光振幅噪声的抑制、激光随机相位抖动产生的入射光相位噪声的抑制、原子噪声的抑制以及相位噪声到振幅噪声转换效率的抑制上。

考虑到里德堡 EIT 系统中耦合光噪声向探测光的转移，本文针对耦合光稳频中的内外调制方式对探测光的输出噪声进行了研究。首先从理论上分析了 EIT 系统中探测光的振幅噪声、探测光的相位噪声、耦合光的振幅噪声、耦合光的相位噪声及原子噪声共同决定了探测光的输出振幅噪声。然后从实验上测量了外调制检测输出端、稳频输出端与内调制检测输出端、稳频输出端的噪声谱，证明了在 EIT 介质中耦合光的调制信号可以向探测光转化，外调制稳频引入的调制信号不会影响检测输出端；较内调制稳频，外调制稳频由于调制所引入的额外相位噪声不会影响检测输出端，有效避免了 EIT 系统检测输出端的振幅噪声。此外，实验观察了相位噪声大小随调制幅度的变化。结果表明，耦合光由于调制所引入的额外相位噪声会给 EIT 系统带来不可忽视的影响。

## 2 基本原理

考虑如图 1 所示的阶梯型三能级系统。基态 $|a\rangle$ 和中间态 $|b\rangle$ 的能级间隔表示为 $\omega_{ab}$，中间态 $|b\rangle$ 和里德堡态 $|c\rangle$ 的能级间隔表示为 $\omega_{bc}$；探测光 $v_p$ 作用于 $|a\rangle \rightarrow |b\rangle$，失谐为 $\triangle_p = \omega_{ab} - v_p$；耦合光 $v_c$ 作用于 $|b\rangle \rightarrow |c\rangle$，失谐为 $\triangle_c = \omega_{bc} - v_c$；双光子失谐表示为 $\delta = \triangle_p + \triangle_c$。

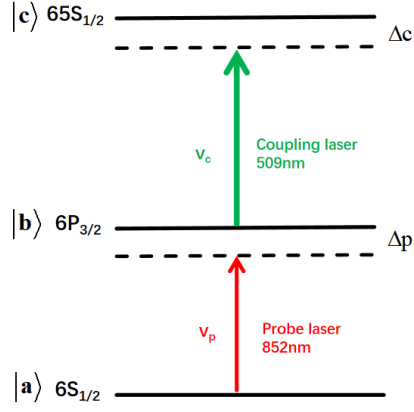

图 1 铯原子阶梯型 EIT 的能级图

Fig. 1 Energy level scheme of the ladder-type EIT of the cesium atom

引入算符：

$$X_A = \frac{1}{2}(a+a^\dagger) \tag{1}$$

$$X_\phi = \frac{1}{2i}(a-a^\dagger) \tag{2}$$

对于量子化的单模电磁场：

$$E(z,t) \equiv 2E^{(s)}\sin(kz)\left[X_A\cos(\omega t)+X_\phi\sin(\omega t)\right] \tag{3}$$

其中 $k$ 是波数，$z$ 是电磁场沿 z 轴方向传播上的距离，$\omega$ 是角频率，$a$ 是湮灭算符，$a^\dagger$ 是产生算符。$X_A$ 对应经典场的振幅涨落，$X_\phi$ 对应经典场的相位涨落。

依据参考文献[16]中提出的方法，输出探测光的正交振幅噪声为：

$$S_A(z,\omega) = S_{A,2}(\omega)+S_{\phi,2}(\omega)+S_{atom}(\omega) \tag{4}$$

$$S_{A,2}(\omega) = \frac{S_{A,2}(0,\omega)}{4}(\exp\{-[\Lambda(\omega)+\Lambda(-\omega)]z\}+\exp\{-[\Lambda(\omega)+\Lambda^*(\omega)]z\} \\ +\exp\{-[\Lambda^*(-\omega)+\Lambda(-\omega)]z\}+\exp\{-[\Lambda^*(-\omega)+\Lambda*(\omega)]z\}) \tag{5}$$

$$S_{\phi,2}(\omega) = \frac{S_{\phi,2}(0,\omega)}{4}(\exp\{-[\Lambda(\omega)+\Lambda(-\omega)]z\}-\exp\{-[\Lambda(\omega)+\Lambda^*(\omega)]z\} \\ -\exp\{-[\Lambda^*(-\omega)+\Lambda(-\omega)]z\}+\exp\{-[\Lambda^*(-\omega)+\Lambda*(\omega)]z\}) \tag{6}$$

$$S_{atom}(\omega) = 1-\exp\left[-2\operatorname{Re}(\Lambda(\omega)z)\right] \tag{7}$$

$$\Lambda(\omega) = \frac{g_2^2 N}{c}\left(\frac{\gamma_0-i(\omega-\delta)}{\left[\gamma-i(\omega-\Delta_p)\right]\left[\gamma_0-i(\omega-\delta)\right]+|\Omega|^2}\right)-\frac{i\omega}{c} \tag{8}$$

其中，$\gamma_0$ 表示基态的退相干率；$\Omega$ 表示耦合光的 Rabi 频率；$g_1$ 代表耦合光与原子的耦合常数；$g_2$ 代表探测光与原子的耦合常数；$N$ 表示原子数密度；$\gamma = \Gamma/2$ 表示 $\sigma_{ca}$ 或 $\sigma_{ba}$ 的衰减率。可见，探测光的输出振幅噪声包括三部分:探测光的输入振幅噪声 $S_{A,2}(\omega)$、探测光的输入相位噪声 $S_{\phi,2}(\omega)$、原子随机衰减的朗之万噪声 $S_{atom}(\omega)$。理论模拟的参数 $S_{A,2}(0,\omega)|_{\omega=1\text{MHz}} = 1$ 代表量子噪声极限，$S_{\phi,2}(0,\omega)|_{\omega=1\text{MHz}} = 38 \text{ dB}$，$\omega$ 代表分析频率，如图 2 所示。EIT 光谱的相位噪声随耦合光失谐变化。

定义电磁场的正交分量的涨落为：

$$\delta X_j^\theta(z,t) = \delta a_j(z,t)\exp(-i\theta) + \delta a_j^\dagger(z,t)\exp(i\theta) \tag{9}$$

其中，$\theta$ 表示单位幅度场的相位，j=1，2 分别表示耦合光和探测光。当 $\theta = 0$，$\delta X_j^0(z,t) = \delta a_j(z,t) + \delta a_j^\dagger(z,t)$，$\delta X$ 对应场的正交振幅涨落；当 $\theta = \pi/2$，$\delta X_j^{\pi/2}(z,t) = i\left[\delta a_j^\dagger(z,t) - \delta a_j(z,t)\right]$，$\delta X$ 对应场的正交相位涨落。

在稳态情况下，正交分量噪声谱由下式给出：

$$S_j(z,\omega) = \int_{-\infty}^{\infty} e^{-i\omega t}\left\langle \delta X_j^\theta(z,t)\delta X_j^\theta(z,0)\right\rangle dt \tag{10}$$

根据参考文献[17]有：

$$S_2(Q^{(r)}z = k\pi) = S_1(Q^{(r)}z = (k+1)\pi) \tag{11}$$

$$S_j(Q^{(r)}z = k\pi + \pi/2) = (S_1(z=0) + S_2(z=0))/2 \tag{12}$$

$$Q^{(r)} = \frac{\omega N(g_1^2\Omega_2^2 + g_2^2\Omega_1^2)}{\Omega^2 c} \times \frac{\Omega^2 - \omega^2}{\left[\Omega^2 - \omega^2\right]^2 + \omega^2\gamma_0^2/4} \tag{13}$$

当光场在介质中传播的相干振荡相位满足 $Q^{(r)}z = k\pi$ 时：$S_2$（探测光）在该位置的噪声等同于 $S_1$（耦合光）在 $Q^{(r)}z = (k+1)\pi$（下一个半周期位置）的噪声。当 $\theta = 0$，耦合光与探测光的振幅噪声发生周期性转移；当 $\theta = \pi/2$，耦合光与探测光的相位噪声发生周期性转移。因此耦合光的振幅噪声 $S_{A,1}(\omega)$、耦合光的相位噪声 $S_{\phi,1}(\omega)$、探测光的振幅噪声 $S_{A,2}(\omega)$、探测光的相位噪声 $S_{\phi,2}(\omega)$、及原子噪声共同决定了探测光的输出振幅噪声。

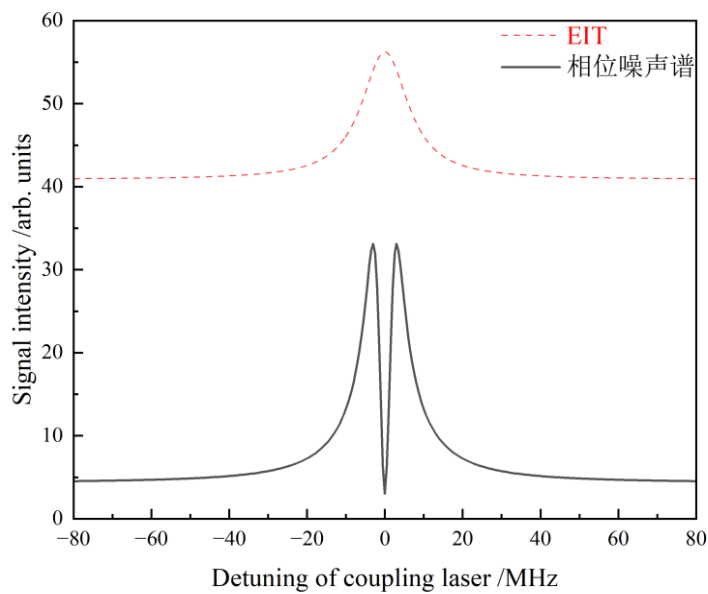

图 2 探测光经 EIT 介质后的噪声转化

Fig. 2 Noise conversion of probe laser by EIT medium

# 3 实验装置与分析讨论

铯原子阶梯型 EIT 的能级图如图 1 所示，波长为 852 nm 的探测光对应基态 $|6S_{1/2}\rangle$ 到中间态 $|6P_{3/2}\rangle$ 的原子跃迁，波长为 509 nm 的耦合光对应中间态 $|6P_{3/2}\rangle$ 到里德堡态 $|65S_{1/2}\rangle$ 的原子跃迁，实现里德堡原子的制备及 EIT 光谱的探测。

噪声谱测量实验装置如图 3（a）所示，利用分束装置将探测光输出光束分成 B1、B2、B3 三束，将耦合光输出光束分为 B4、B5 两束。其中，光束 B1 被用来实现探测光稳频。光束 B2 与 B4 共线反向经过铯原子气室用于搭建 EIT 光谱，实现耦合光外调制稳频或内调制稳频，并进行稳频输出端噪声谱的测量，如图 3（a）中虚线框图所示。光束 B3 与 B5 共线反向经过铯原子气室用于搭建 EIT 光谱，进行检测输出端噪声谱的测量，如图 3（a）中实线框图所示。图 3（b）、3（c）所示为图 3（a）中虚线框图的具体实验装置图，用于实现耦合光稳频。其中图 3（b）为耦合光外调制稳频示意图，信号发生器输出正弦波信号给 EOM，信号频率为 10.3 MHz，幅度为 1.2 V。耦合光 B4 经过 EOM 进行相位调制后进入铯原子气室，与探测光 B2 在铯原子气室中反向共线传输。两束光在原子气室中发生非线性四波混频，使耦合光上的调制信号转移到探测光上，探测光经过双色镜 DM3 进入光电探测器 2，将光电探测器 2 转换后的电信号与相移器信号进行混频，再经过低通滤波器将其中的高

频部分与直流信号滤除，最后通过示波器可以观察到铯原子 D2 线的鉴频信号。得到的鉴频信号输入到 PID 后驱动压电陶瓷，在不扫描耦合光的情况下通过鉴频信号实现外调制稳频。图 3（c）为耦合光内调制稳频示意图，信号发生器输出频率为 52.5 kHz、幅度为 0.004 V 的正弦信号对耦合光激光器的压电陶瓷进行调制，带调制的耦合光 B4 进入铯原子气室，与探测光 B2 在铯原子气室中反向共线传输。两束光在原子气室中发生非线性四波混频，使耦合光上的调制信号转移到探测光上，将光电探测器 2 转换后的电信号与相移器信号进行混频，再经过低通滤波器将其中的高频部分与直流信号滤除，将鉴频信号输入 PID 后驱动压电陶瓷，在不扫描耦合光的情况下通过鉴频信号实现内调制稳频。

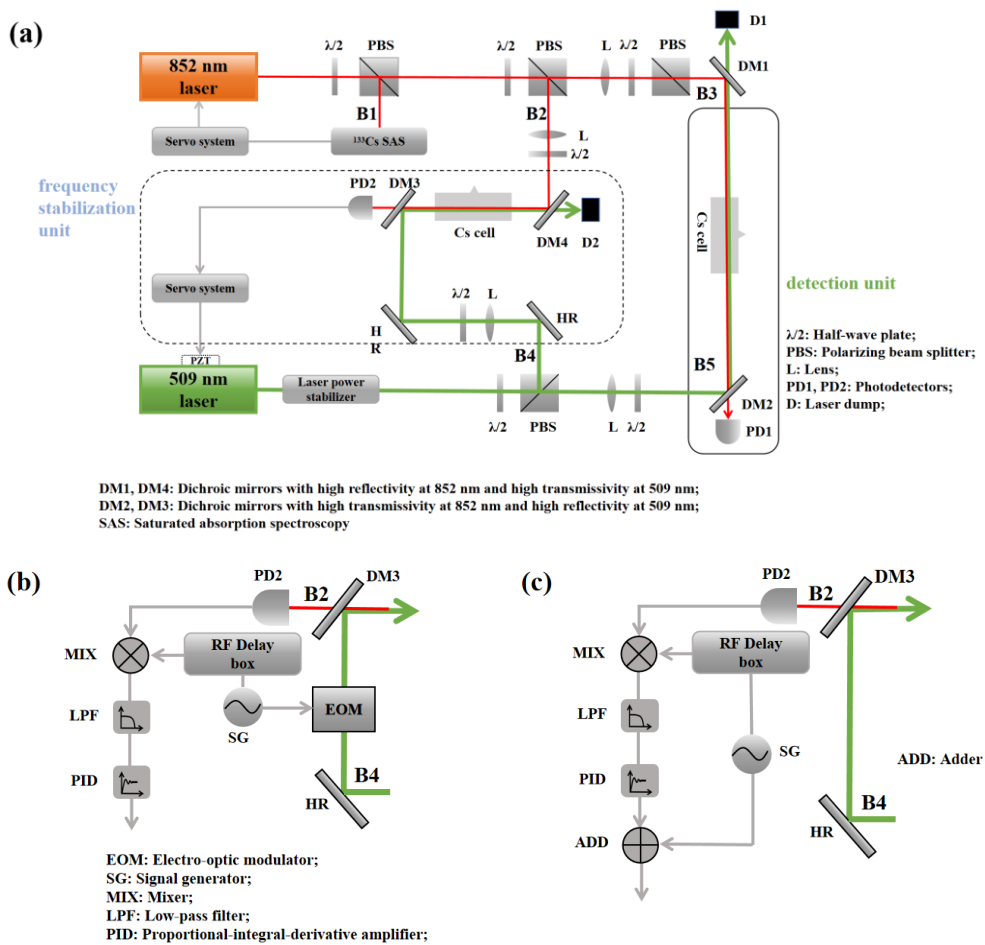

图 3 实验装置图。（a）噪声谱测量实验装置图。（b）外调制稳频示意图。（c）内调制稳频示意图。

Fig. 3 Schematic diagram of experimental setup: (a) Schematic of noise spectrum measurement setup. (b) External modulation frequency stabilization schematic. (c) Internal modulation frequency stabilization schematic.

频谱分析仪将时域中随机的噪声信号转换为频域中的功率谱密度进行分析，从而量化噪声在不同频率上的分布特征。我们将 PD 转化后的电信号接入频谱分析仪，在双光子共振时测量了稳频输出端与检测输出端的噪声谱。如图 4（a）所示为外调制稳频输出端与检测输出端的噪声谱，设置频谱分析仪的视频带宽为 100 Hz，分析带宽为 1 kHz，起始频率为 0 Hz，

终止频率为 30 MHz。在 EOM 的作用下，外调制稳频输出端探测光被引入了约 5 dB、10.3 MHz 的调制信号；如图 4（b）所示为内调制稳频输出端与检测输出端的噪声谱，设置频谱分析仪视频带宽为 100 Hz，分析带宽为 1 kHz，起始频率为 0 Hz，终止频率为 200 kHz。在锁相放大器作用下，内调制稳频输出端及检测输出端探测光被引入了约 20 dB、52.5 kHz 的调制信号。因此在 EIT 介质中耦合光的调制信号可以向探测光转化。如图（a）实线所示，外调制检测输出端探测光未引入调制信号，因此外调制稳频引入的调制信号不会影响检测输出端。此外可以看到，出现调制信号转化的噪声谱线除观察到所加的调制信号外，还观察到了调制信号的二倍频信号，这是由于调制器的非线性效应。

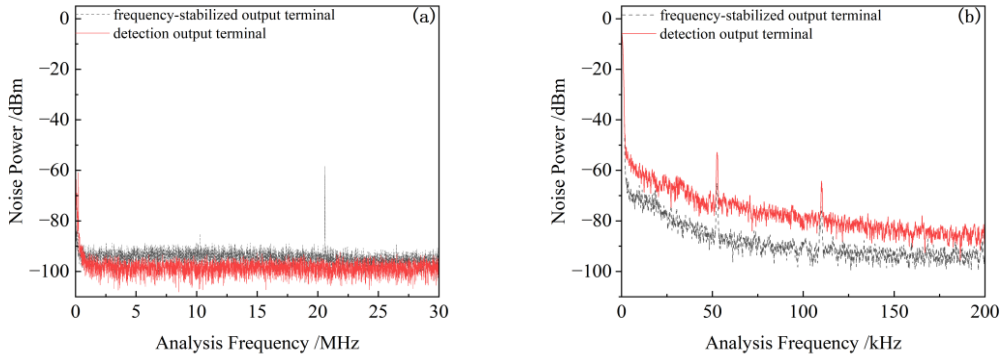

图 4 调制信号从耦合光到探测光的转化。（a）外调制稳频；（b）内调制稳频

Fig. 4 The conversion of the modulated signal from the coupling light to the probe light. (a) External modulation frequency-stabilization; (b) Internal modulation frequency-stabilization

实验系统的探测光未进行相位锁定，输出探测光的噪声谱不仅包括振幅噪声还包括相位噪声，下面我们对实验中的噪声贡献进行分析。我们采用北京无线电计量测试研究所生产的 MI-10450-45-9 功率稳定仪对耦合光激光器进行功率稳定，其输出光强的随机涨落对探测光输出振幅噪声 $S_A$ 的贡献 $S_{A,1}$ 忽略不计。探测光激光器未进行功率稳定，其输出光强的随机涨落对 $S_A$ 的贡献为 $S_{A,2}$。耦合光激光器和探测光激光器未进行相位锁定，其随机相位抖动对 $S_A$ 的贡献分别为 $S_{\phi,1}$、$S_{\phi,2}$。耦合光由于稳频所加的调制信号对耦合光相位噪声的贡献为 $S_{\phi',1}$。探测光与耦合光之间未进行相位锁定，当 $\theta = \pi/2$ 时，在原子气室中耦合光相位噪声 $S_{\phi,1}$、$S_{\phi',1}$ 向探测光相位噪声的转移对 $S_A$ 的贡献为 $\varepsilon_{\phi,1}S_{\phi,1}$、$\varepsilon_{\phi',1}S_{\phi',1}$。$S_{A,2}$、$\varepsilon_{\phi,1}S_{\phi,1}$、$S_{\phi,2}$、$\varepsilon_{\phi',1}S_{\phi',1}$ 及原子随机衰减过程产生的朗之万噪声 $S_{atom}$ 共同决定了探测光的输出振幅噪声 $S_A$：

$$S_A = \sqrt{S_{A,2}^2 + \varepsilon_{\phi,1}^2 S_{\phi,1}^2 + S_{\phi,2}^2 + \varepsilon_{\phi',1}^2 S_{\phi',1}^2 + S_{atom}^2} \tag{14}$$

EIT 光谱的相位噪声随耦合光失谐变化，将 PD 转化后的电信号接入频谱分析仪，在耦

合光激光器的扫描频率为 0.2 Hz 时测量稳频输出端与检测输出端的噪声谱，如图 5 所示，设置频谱分析仪为零扫宽模式，视频带宽为 5.1 Hz，分析带宽为 5.1 Hz。我们知道此时的系统中，稳频输出端及检测输出端的探测光激光器与耦合光激光器相同，$S_{A,2}$、$\varepsilon_{\phi,1}S_{\phi,1}$、$S_{\phi,2}$、$S_{atom}$ 对稳频输出端及检测输出端的噪声谱贡献相同。由图 5 可知，外调制检测输出端的噪声谱接近一条水平直线，因此入射光的振幅噪声 $S_{A,2}$、激光器自身相位抖动产生的相位噪声 $\varepsilon_{\phi,1}S_{\phi,1}$、$S_{\phi,2}$ 以及原子随机衰减过程产生的朗之万噪声 $S_{atom}$ 对"M"型噪声增强现象的贡献可以忽略不计，外调制稳频输出端、内调制稳频输出端以及内调制检测输出端的噪声增强主要来自耦合光稳频所引入的额外相位噪声 $S_{\phi',1}$ 到输出振幅噪声的转化 $\varepsilon_{\phi',1}S_{\phi',1}$。探测光输出振幅噪声 $S_A$ 的噪声谱的噪声增强部分 $S'_A$：

$$S'_A = \sqrt{\varepsilon^2_{\phi',1}S^2_{\phi',1}} \tag{15}$$

外调制检测输出端的噪声谱接近一条水平直线，无额外相位噪声转移，因此外调制稳频未给激光器引入额外的相位噪声。内调制检测输出端的噪声谱呈"M"型，在±7.4 MHz 的频率失谐处，光相位噪声转化达到最大值，约 39 dB；当探测光与耦合光的频率满足双光子共振条件时，相位噪声转化约 30 dB。因此外调制稳频相比内调制稳频，有效避免了 EIT 系统检测输出端的振幅噪声。在±7.4 MHz 的频率失谐处，内调制稳频输出端耦合光相位噪声转化达到最大值，与外调制稳频输出端一致；在±10.6 MHz 的频率失谐处，内调制检测输出端的耦合光相位噪声转化达到最大值，区别于两种稳频方法的稳频输出端。两种方法的稳频输出端的 EIT 光谱来自同一套 EIT 系统，如图 3（a）中虚线框图所示，具有相同的 EIT 线宽。而内调制检测输出端采用另一套 EIT 系统，如图 3（a）中实线框图所示，其 EIT 线宽与前者不同。考虑到 EIT 线宽越窄，原子相干时间越长。在相位噪声转化过程中，原子与光场相互作用时间越长，相位噪声转化为输出振幅噪声概率更大。因此噪声转化最大值对应的频率失谐范围可能与 EIT 线宽有关。

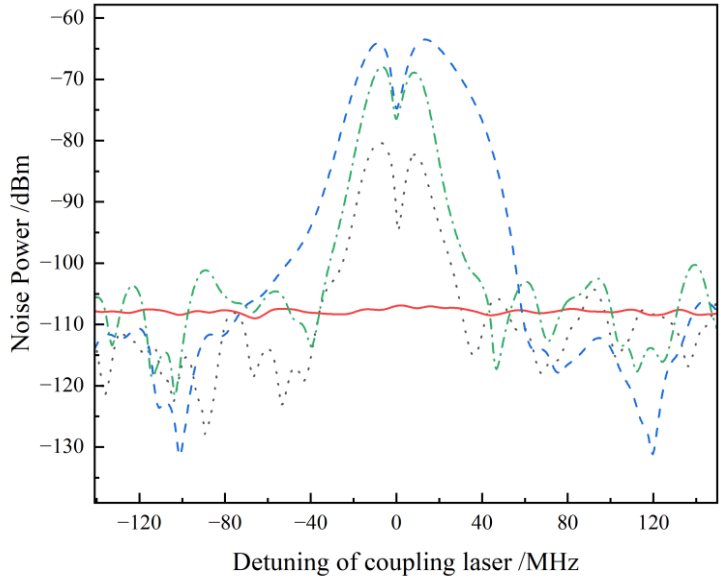

图 5 探测光噪声谱。点线和实线分别代表外调制稳频条件下稳频输出端与检测输出端的噪声谱；划线和点划线分别代表内调制稳频条件下稳频输出端与检测输出端的噪声谱

Fig. 5 The noise spectrum of the probe laser. The dotted lines and solid line represent the noise spectra of the requency-stabilized output terminal and detection output terminal under external modulation frequency-stabilization conditions, respectively. The lineation and dot-dash line represent the noise spectra of the frequency-stabilized output terminal and detection output terminal under internal modulation frequency-stabilization conditions, respectively

改变耦合光所加调制信号的调制幅度，研究额外相位噪声大小与调制幅度的依赖关系，结果如图 6 所示。外调制稳频的调制幅度在 0.2~2.0 V 的变化范围内，耦合光的额外相位噪声增加了约 10 dB；内调制稳频的调制幅度在 0.004~0.022 V 的变化范围内，耦合光的额外相位噪声大小增加了约 15 dB。耦合光由于调制所引入的额外相位噪声会给 EIT 系统带来不可忽视的影响。因此在 EIT 系统中采用外调制方法实现耦合光稳频，有利于抑制检测输出端由于稳频所引入的额外相位噪声。

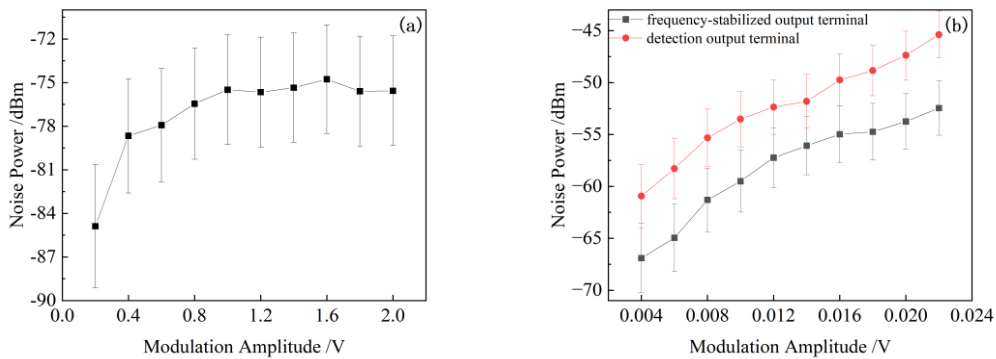

图 6 耦合光额外相位噪声随调制幅度的变化。（a）外调制稳频；（b）内调制稳频

Fig. 6 The variation of coupling laser excess phase noise with modulation amplitude. (a) External modulation frequency-stabilization; (b) Internal modulation frequency-stabilization

## 4 结　论

本文研究了外调制稳频与内调制稳频里德堡 EIT 系统耦合光调制信号及耦合光额外相位噪声的转化，证明了外调制稳频可有效避免调制信号和调制噪声转移。理论分析得到了探测光的振幅噪声、探测光的相位噪声、耦合光的振幅噪声、耦合光的相位噪声及原子噪声共同决定探测光的输出振幅噪声。在双光子共振时测量了外调制稳频与内调制稳频下，稳频输出端与检测输出端的噪声谱，实验观察到在 EIT 介质中耦合光的调制信号可以向探测光转化，外调制稳频引入的调制信号不会影响检测输出端，内调制的检测输出端在调制幅度为 0.004 V 时被引入了约 20 dB 的调制信号。在频谱分析仪零扫宽设置时测量了外调制稳频与内调制稳频下，稳频输出端与检测输出端的噪声谱。外调制检测输出端的噪声谱接近一条水平直线，输出探测光的噪声谱主要来自耦合光的额外相位噪声到振幅噪声的转化。内调制检测输出端的噪声谱呈"M"型，在±7.4 MHz 的频率失谐处，光相位噪声转化达到最大值，约 39 dB；当探测光与耦合光的频率满足双光子共振条件时，相位噪声转化约 30 dB。因此外调制稳频有效避免了 EIT 系统检测输出端的振幅噪声。转化噪声转化最大值对应的频率失谐范围可能与 EIT 线宽有关。此外，实验观察了外调制稳频输出端相位噪声大小随调制幅度的变化以及内调制稳频输出端与检测输出端相位噪声大小随调制幅度的变化。结果表明，耦合光由于调制所引入的额外相位噪声会给 EIT 系统带来不可忽视的影响，在基于电磁诱导透明效应进行测量和通信实验时，采用外调制方案实现耦合光稳频，有利于抑制检测输出端由于稳频所引入的额外相位噪声。在噪声不是实验的主要影响因素或对实验的测量精度没有更高的要求时采用内调制更为便捷。

## 参考文献

# Influence of Modulation Frequency Stabilization on Spectral Noise of Electromagnetically Induced Transparency

Hou Jinghua[1], Su Nan[2], Liu Yao[2], Liu Zhihui[2], Zhang Yuchi[1]*, He Jun[2]**

[1] *School of Physics and Electronic Engineering, Shanxi University, Taiyuan, Shanxi 030006, China*;

[2] *State Key Laboratory of Quantum Optics and Quantum Optics Devices, Institute of Opto-Electronics, Shanxi University, Taiyuan, Shanxi 030006, China*

**Abstract**

**Objective** The electromagnetically induced transparency effect is a typical nonlinear coherent optical phenomenon. In a three-level atomic system, a strong coupling light and a weak probe light act together: through quantum interference, the absorption of the probe light by atoms is suppressed, thereby rendering the originally opaque medium transparent. The electromagnetically induced transparency effect has not only been realized in atomic vapor cells but also extensively studied in media such as cold atoms, solid-state materials, and semiconductors. The application of electromagnetically induced transparency technology in atomic vapor cells or cold atoms has enabled microwave electric field measurement and communication based on Rydberg atoms, driving rapid development in these fields. In measurement and communication systems based on atomic vapor cells, noise is one of the key factors affecting measurement sensitivity and communication bit error rate. Consequently, extensive research has been conducted on the noise characteristics of electromagnetically induced transparency signals. Studies have shown that the output amplitude noise of the probe light in an electromagnetically induced transparency system is collectively determined by five components: the amplitude noise of the probe light, the phase noise of the probe light, the amplitude noise of the coupling light, the phase noise of the coupling light, and atomic noise. To reduce the noise of the output probe light, efforts have been made from multiple perspectives. Noise suppression work mainly focuses on four aspects: suppressing the incident light amplitude noise caused by laser power fluctuations, suppressing the incident light phase noise induced by random laser phase jitter, suppressing atomic noise, and suppressing the conversion efficiency of phase noise to amplitude noise.

**Methods** Through theoretical analysis, this paper concludes that the output amplitude noise of the probe light is collectively determined by five components: the amplitude noise of the probe light, the phase noise of the probe light, the amplitude noise of the coupling light, the phase noise of the coupling light, and atomic noise. In the experiment, two frequency stabilization methods—external modulation transfer spectroscopy and internal modulation transfer spectroscopy—were employed respectively to achieve stable frequency of the coupling light. Furthermore, under the conditions of two-photon resonance and zero-span setting of the spectrum analyzer, the probe light noise spectra at the frequency-stabilized output and detection output were measured for both frequency stabilization methods.

**Results and Discussions** When two-photon resonance was achieved, the noise spectra at the frequency-stabilized output and detection output were measured for both external modulation-based and internal modulation-based frequency stabilization. Experimental observations revealed that the modulation signal of the coupling light in the electromagnetically induced transparency medium can be converted to the probe light: the modulation signal introduced by external modulation-based frequency stabilization did not affect the detection output, whereas for internal modulation, a modulation signal of approximately 20 dB was introduced at the detection output when the modulation amplitude was 0.004 V (Fig.4). Moreover, the noise spectra at the frequency-stabilized output and detection output were measured under the zero-span setting of the spectrum analyzer for both frequency stabilization methods. The noise spectrum at the external modulation detection output was close to a horizontal line, indicating that the noise spectrum of the output probe light mainly originated from the conversion of additional phase noise of the coupling light to amplitude noise. In contrast, the noise spectrum at the internal modulation detection output exhibited an "M" shape: the conversion of optical phase noise reached a maximum of approximately 39 dB at a frequency detuning of $\pm 7.4$ MHz, and approximately 30 dB when the frequencies of the probe light and coupling light satisfied the two-photon resonance condition. Thus, external modulation-based frequency stabilization effectively avoids amplitude noise at the detection output of the electromagnetically induced transparency system. The frequency detuning range corresponding to the maximum noise conversion may be related to the electromagnetically induced transparency linewidth (Fig.5). Additionally, the experiment observed the variation of phase noise at the frequency-stabilized output with modulation amplitude for external modulation, as well as the variation of phase noise at both the frequency-stabilized output and detection output with modulation amplitude for internal modulation. When the modulation amplitude of external modulation-based frequency stabilization varied in the range of 0.2–2.0 V, the additional phase noise of the coupling light increased by approximately 10 dB; when the modulation amplitude of internal modulation-based frequency stabilization varied in the range of 0.004–0.022 V, the additional phase noise of the coupling light increased by approximately 15 dB (Fig.6).

**Conclusions** The results indicate that the additional phase noise introduced by the modulation of the coupling light exerts a non-negligible impact on the electromagnetically induced transparency system. Therefore, in measurement and communication experiments based on the electromagnetically induced transparency effect, the external modulation scheme can be adopted to achieve frequency stabilization of the coupling light, thereby avoiding the additional phase noise introduced by frequency stabilization at the detection output.

**Key words**　noise spectrum; electromagnetically induced transparency; laser frequency stabilization; excess phase noise